\documentclass[aps,prl,preprint]{revtex4-2} 

\usepackage{amsmath}  
\usepackage{amssymb}
\usepackage{mathrsfs}
\usepackage{amsfonts} 
\usepackage{esvect}
\usepackage{graphicx} 
\usepackage{hyperref}
    \hypersetup{colorlinks,urlcolor=blue,linkcolor=blue,citecolor=blue}
\usepackage{mhchem}
\usepackage{physics}

\usepackage{floatrow}
\usepackage{subfig}

\newcommand*{\tensortwo}[1]{\bar{\bar{#1}}}

\begin{document}
\title{Scalable cell filter nudged elastic band (CFNEB) for large-scale transition-path calculations}
\author{Qiuhan Jia}
\affiliation{School of Physics, Nanjing University, Nanjing, P.R. China, 210093}

\author{Jiuyang Shi}
\affiliation{School of Physics, Nanjing University, Nanjing, P.R. China, 210093}

\author{Jian Sun}
\affiliation{School of Physics, Nanjing University, Nanjing, P.R. China, 210093}

\date{\today}

\begin{abstract}
The nudged elastic band (NEB) method is one of the most widely used techniques for determining minimum-energy reaction pathways and activation barriers between known initial and final states. However, conventional implementations face steep computational scaling with system size, which makes nucleation-type transitions in realistically large supercells practically inaccessible. In this work, we develop a scalable cell-filter nudged elastic band (CFNEB) framework that enables efficient transition-path calculations in systems containing up to $10^5$ atoms. The method combines a deformation-based cell filtering scheme, which treats lattice vectors as generalized coordinates while removing spurious rotational degrees of freedom, with an adaptive image insertion and deletion strategy that dynamically refines the reaction path. We implement CFNEB both within the ASE environment and in a fully GPU-accelerated version using the Graphics Processing Units Molecular Dynamics (GPUMD) engine, achieving throughput on the order of $10^6$ atom$\cdot$steps per second on consumer GPUs. We demonstrate the method on two representative systems: the layer-by-layer $\beta$-$\lambda$ transition in \ce{Ti3O5} and the nucleation-driven graphite-to-diamond transformation. These examples illustrate that CFNEB not only reproduces known concerted pathways but also captures spontaneous symmetry breaking toward nucleated mechanisms when the simulation cell is sufficiently large. Our results establish CFNEB as a practical route to exploring realistic transition mechanisms in large-scale solid-state systems.
\end{abstract}

\maketitle

\floatsetup[figure]{style=plain,subcapbesideposition=top}
\section{Introduction}
Transitions in condensed phases—whether chemical reactions, diffusion events, or solid–solid transformations—are governed by path on a high-dimensional energy landscape connecting long-lived metastable states. Within the framework of Transition State Theory (TST), the system’s rate of transformation is determined by the probability of crossing a dividing surface that passes through the saddle region of the potential energy surface (PES), corresponding to the transition state.\cite{eyring_activated_1935,mills_generalized_1997,laidler_development_1983,bao_variational_2017} A transition path specifies how the system leaves a basin near the initial state, crosses a saddle region (the rate-limiting bottleneck), and enters the basin of the final state. On the zero-temperature potential energy surface (PES), the most relevant object is the minimum-energy path (MEP), whose highest point determines the activation energy barrier. At finite temperature, free-energy generalizations of the path become important, but the PES MEP remains a central reference for mechanism identification, saddle search, and for constructing physically meaningful reaction coordinates that are compatible with transition-state theory rates.

In crystalline solids, the transition path generally involves coupled evolution of atomic positions and lattice degrees of freedom, as local rearrangements can be accompanied by strain relaxation, shear, or volume change\cite{sheppard_generalized_2012,zhang_variablecell_2015}. Moreover, the physically realized mechanism often depends on system size and boundary conditions: while small supercells tend to favor concerted or homogeneous distortions, realistic transformations commonly proceed via localized nucleation-and-growth with long-range elastic fields\cite{mahata_size_2019, santos-florez_sizedependent_2022,zhang_morphology_2007,zarkevich_lattice_2018}. A computational framework for transition paths must therefore (i) represent a continuous path by a tractable discrete object, (ii) robustly locate saddle regions without sliding off the reaction coordinate, and (iii) remain effective as system size grows, where heterogeneous pathways emerge.

The nudged elastic band (NEB) method provides such a representation by discretizing the path into a chain of images and relaxing it toward the MEP using force projections and elastic springs\cite{jonsson_nudged_1998,henkelman_climbing_2000,henkelman_improved_2000}. Since its original formulation in the late 1990s\cite{jonsson_nudged_1998}, refinements such as the improved tangent estimates\cite{henkelman_improved_2000} and climbing-image NEB (CI-NEB)\cite{henkelman_climbing_2000} have stabilized path optimization and accelerated saddle convergence. Variable-cell extensions—including VCNEB \cite{qian_variable_2013} and G-SSNEB \cite{sheppard_generalized_2012}—incorporate lattice degrees of freedom to address pressure- or stress-driven transitions, thereby broadening the scope of solid-state applications. More recently, finite deformation nudged elastic band (FD-NEB) has been developed to solve the problem of non-hydrostatic pressure\cite{ghasemi_method_2020,ghasemi_nudged_2024}.

Despite these advances, computational cost grows with both image count and system size, which effectively limits most NEB studies to supercells of a few hundred atoms and biases reported mechanisms toward concerted distortions. As a result, nucleation phenomena that require large dimensions are frequently inaccessible. To address this, we develop a cell-filter nudged elastic band (CFNEB) framework aimed explicitly at scalable transition-path sampling in large systems. Our approach combines a deformation-based generalized coordinate transformation that couples atomic and lattice relaxation while eliminating inconsequential rotational modes, with an automatic image insertion/pruning algorithm that adaptively controls the intervals of replicas along the path. Most importantly, we implement CFNEB within a fully GPU-accelerated infrastructure (GPUMD), enabling simulations with up to 100 images and $10^5$ atoms at practical wall times.

\section{Theory}
\subsection{NEB}
In NEB, the transition path between two known endpoint configurations is represented by a discrete chain of intermediate images. Each image is subject to two forces: a real force from the underlying potential energy surface and a spring force that enforces path continuity. To prevent unwanted sliding along the reaction coordinate, only the component of the real force perpendicular to the local tangent is retained, while the spring force is projected along the tangent:  
\begin{align}
    f^{\mathrm{NEB}} = \left.f^{\mathrm{real}}\right|_\perp + f^{\mathrm{spring}}.
\end{align}
This projection scheme ensures that relaxation occurs toward the MEP while maintaining reasonable spacing between images.

\subsection{Stress and virial}
Using the right-multiplication definition of strain, the deformed lattice matrix is
\begin{equation}\label{epsilon}
    \mathbf{h}' = \mathbf{h}\,(\mathbf{I} + \boldsymbol{\varepsilon}),
\end{equation}
where \(\mathbf{h} = (\mathbf{h}_1,\mathbf{h}_2,\mathbf{h}_3)^T\) collects the lattice row vectors \(\mathbf{h}_i\), \(\mathbf{h}'\) is the deformed lattice, and \(\tensortwo{\varepsilon}\) is the (infinitesimal) strain tensor. The virial is the negative derivative of the total energy with respect to strain,
\begin{equation}\label{V}
    \tensortwo{V} = -\,\frac{\partial U}{\partial \tensortwo{\varepsilon}},
\end{equation}
with \(U\) the total potential energy. Assume \(U\) can be decomposed over atoms as
\begin{equation}\label{U}
    U = \sum_i U_i = \sum_i U_i\!\big(\mathbf{r}^{i,i_1},\ldots,\mathbf{r}^{i,i_j},\ldots,\mathbf{r}^{i,i_{n_i}}\big),
\end{equation}
where \(\mathbf{r}^{ij} \equiv \mathbf{r}^j - \mathbf{r}^i\) is the relative displacement between atoms \(i\) and \(j\), and \(n_i\) counts neighbors in \(i\)'s local environment under PBC. The interatomic force on \(i\) due to \(j\) is
\begin{equation}
    \mathbf{f}^{ij} \;=\; \frac{\partial U_j}{\partial \mathbf{r}^{ij}} \;-\; \frac{\partial U_i}{\partial \mathbf{r}^{ji}} .
\end{equation}

Substituting Eq.~\eqref{U} into Eq.~\eqref{V} and using Einstein notation gives
\begin{align}\label{V2}
    V_{mn}
    &= -\frac{\partial U}{\partial \varepsilon_{mn}}
     = -\sum_i \frac{\partial U_i}{\partial \varepsilon_{mn}}
     = -\sum_i \sum_j \frac{\partial U_i}{\partial r^{ij}_k}\,\frac{\partial r^{ij}_k}{\partial \varepsilon_{mn}}
     \nonumber\\
    &= \sum_{i,j} \frac{\partial U_i}{\partial r^{ij}_k}\, s^{ij}_\ell \, \frac{\partial h_{\ell k}}{\partial \varepsilon_{mn}},
\end{align}
where \(\mathbf{s}\) denotes fractional coordinates and \(s^{ij} \equiv \mathbf{s}^j - \mathbf{s}^i\). From Eq.~\eqref{epsilon},
\(\Delta h_{\ell k} = h_{\ell p}\,\varepsilon_{pk} = h_{\ell p}\,\varepsilon_{mn}\delta_{nk}\), hence \(\partial h_{\ell k}/\partial \varepsilon_{mn} = h_{\ell m}\delta_{nk}\). Therefore,
\begin{align}
    V_{mn}
    &= \sum_{i,j} \frac{\partial U_i}{\partial r^{ij}_k}\, s^{ij}_\ell \, h_{\ell m}\delta_{nk}
     \;=\; -\sum_{i,j} \frac{\partial U_i}{\partial r^{ij}_n}\, r^{ij}_m
    \nonumber\\
    &= -\tfrac{1}{2}\sum_{i,j}\!\left(
          \frac{\partial U_i}{\partial r^{ij}_n}\, r^{ij}_m
        + \frac{\partial U_j}{\partial r^{\,ji}_n}\, r^{ji}_m
       \right) 
    = -\tfrac{1}{2}\sum_{i,j}\!\left(
          \frac{\partial U_i}{\partial r^{ij}_n}
        - \frac{\partial U_j}{\partial r^{\,ji}_n}
       \right) r^{ij}_m
    \nonumber\\
    &= -\tfrac{1}{2}\sum_{i,j} f^{ij}_n\, r^{ij}_m .
\end{align}
In compact vector form,
\begin{equation}
    \mathbf{V} \;=\; -\frac{1}{2}\sum_{i,j} \mathbf{r}^{ij} \otimes \mathbf{f}^{ij}.
\end{equation}
Finally, the Cauchy stress \(\boldsymbol{\sigma}\) follows by dividing the virial by the cell volume \(\Omega\)\cite{fan_force_2015},
\begin{equation}
    \boldsymbol{\sigma} \;=\; \mathbf{V}/\Omega \;=\; -\frac{1}{2\Omega}\sum_{i,j} \mathbf{r}^{ij} \otimes \mathbf{f}^{ij}.
\end{equation}

\subsection{Cell Filter}
The cell filter provides a unified way to relax lattice vectors together with atomic positions during structure optimization \cite{tadmor_mixed_1999}. We concatenate the three lattice vectors with the $n$ position vectors to form a set of generalized positions, and likewise concatenate the three stress vectors with the $n$ atomic forces to form the generalized forces. When lattice and position degrees of freedom are treated as independent, these $(n+3)$ positions can be updated directly by the corresponding $(n+3)$ forces, allowing the wrapped system to be optimized as if it contained $n+3$ “atoms” using standard optimizers.

A naive choice of generalized variables is the pair (fractional coordinates, lattice vectors). Although formally valid, it is numerically inconvenient: fractional coordinates and lattice vectors differ widely in magnitude, and even fractional components can differ significantly in highly non-cubic cells. To obtain variables with comparable scales that remain consistent as system size changes, we work with a reference cell and define the deformation gradient $\mathbf{D} = \mathbf{h}_0^{-1}\,\mathbf{h},$ where 
Atomic positions are recast in a modified basis
\begin{equation}
    \mathbf{\Tilde{R}} = \mathbf{s}\,\mathbf{h}_0
    = \mathbf{R}\,\mathbf{h}^{-1}\mathbf{h}_0
    = \mathbf{R}\,\mathbf{D}^{-1},
\end{equation}
where $\mathbf{\Tilde{R}}$ and $\mathbf{s}$ are $n\times3$ matrices. The generalized coordinates are then
\begin{equation}
    \mathcal{R} = \big(\Tilde{R}_1,\ldots,\Tilde{R}_n,\, D_1,D_2,D_3\big)^T .
\end{equation}

To obtain the conjugate generalized forces, we use virtual work. For the atomic part,
\begin{equation}
    \left.-\Delta E\right|_{\varepsilon} = \mathbf{F} : \Delta\mathbf{R}
    = \mathrm{Tr}\!\left[\mathbf{F}\,\Delta\mathbf{R}^{\,T}\right]
    = \mathrm{Tr}\!\left[\mathbf{F}\,\big(\Delta\mathbf{\Tilde{R}}\,\mathbf{D}\big)^{\!T}\right]
    = (\mathbf{F}\,\mathbf{D}^T):\Delta\mathbf{\Tilde{R}},
\end{equation}
so the modified force that couples to $\Tilde{R}$ is $\mathbf{F}\,\mathbf{D}^T$. For the lattice part,
\begin{equation}
    \left.-\Delta E\right|_{s} = \mathbf{V}:\mathbf{\varepsilon}
    = \mathrm{Tr}\!\left[\mathbf{V}\,\mathbf{\varepsilon}^{\,T}\right]
    = \mathrm{Tr}\!\left[\mathbf{V}\,\big(\mathbf{D}^{-1}\Delta\mathbf{D}\big)^{\!T}\right]
    = \mathrm{Tr}\!\left[\mathbf{D}^{-T}\mathbf{V}\,\Delta\mathbf{D}^{\,T}\right],
\end{equation}
where we used $\mathbf{\varepsilon}=\mathbf{h}^{-1}\Delta\mathbf{h}=(\mathbf{h}_0\mathbf{D})^{-1}\mathbf{h}_0\Delta\mathbf{D}=\mathbf{D}^{-1}\Delta\mathbf{D}$. Hence, the quantity conjugate to $\mathbf{D}$ is $\mathbf{D}^{-T}\mathbf{V}$. Collecting terms, the generalized force vector reads
\begin{equation}
    \mathcal{F} = \big(F_1\mathbf{D}^T,\ldots,F_n\mathbf{D}^T,\, \mathbf{D}^{-T}\!V_1, \mathbf{D}^{-T}\!V_2, \mathbf{D}^{-T}\!V_3\big)^T .
\end{equation}
This cell filter decouples position and lattice updates while keeping the magnitudes of the new coordinates well conditioned; because $\mathbf{D}$ is a deformation measure relative to $\mathbf{h}_0$, consistency under system-size changes is also maintained.

The structure of the modified lattice force resembles the first Piola–Kirchhoff stress.\cite{bonet_nonlinear_1997} Motivated by this, we construct a rotation-free variant by switching to a symmetric strain measure according to the concept of the second Piola–Kirchhoff stress tensors. Define the (left) Cauchy–Green strain tensor
\begin{equation}
    \mathbf{B} = \mathbf{D}\,\mathbf{D}^T ,
\end{equation}
which is invariant under rigid rotations of the lattice ($\mathbf{h}'=\mathbf{h}\mathbf{Q}$ with unitary $\mathbf{Q}$). Rewriting the lattice virtual work in terms of $\mathbf{B}$,
\begin{align}
    \left.-\Delta E\right|_{s}
    &= \mathbf{V}:\mathbf{\varepsilon}
     = \mathrm{Tr}\!\left[\mathbf{D}^{-T}\mathbf{V}\,\Delta\mathbf{D}^{\,T}\right]
     = \big(\mathbf{D}^{-T}\mathbf{V}\,\Delta\mathbf{D}^{-1}\big):\big(\mathbf{D}\,\Delta\mathbf{D}^{\,T}\big) \nonumber\\
    &= \big(\mathbf{D}^{-T}\mathbf{V}\,\Delta\mathbf{D}^{-1}\big) : \frac{1}{2}\!\left(\mathbf{D}\,\Delta\mathbf{D}^{\,T} + \Delta\mathbf{D}\,\mathbf{D}^{\,T}\right)
     = \big(\mathbf{D}^{-T}\mathbf{V}\,\Delta\mathbf{D}^{-1}\big) : \Delta\mathbf{B},
\end{align}
so the generalized lattice force conjugate to $\mathbf{B}$ is $\mathbf{D}^{-T}\mathbf{V}\,\mathbf{D}^{-1}$. The final rotation-free generalized variables and forces are
\begin{equation}
    \mathcal{R} = \big(\Tilde{R}_1,\ldots,\Tilde{R}_n,\, \mathbf{D}\mathbf{D}^{T}\big)^T, 
    \qquad
    \mathcal{F} = \big(F_1\mathbf{D}^{T},\ldots,F_n\mathbf{D}^{T},\, \mathbf{D}^{-T}\mathbf{V}\,\mathbf{D}^{-1}\big)^T .
\end{equation}
Because $\mathbf{B}$ is symmetric (six independent components), the three rotational degrees of freedom are eliminated by construction, yielding a numerically stable, rotation-free cell filter suitable for NEB with large deformations or anisotropic stress.

\section{Implementation}
\begin{figure}[t]
    \begin{center}
    \includegraphics[height=0.4\textwidth]{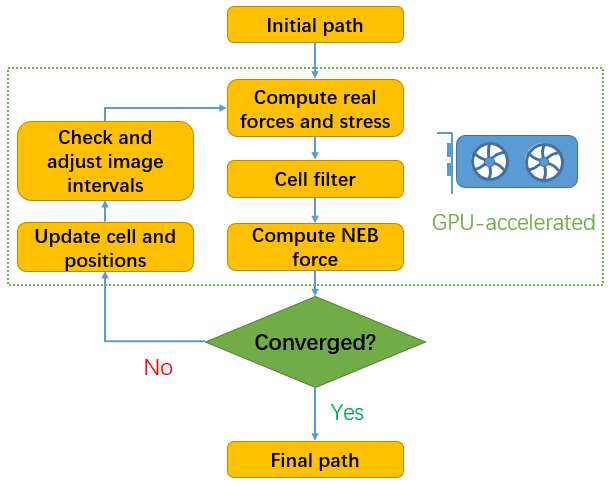}
    \end{center}
    \caption{The workflow diagram of CFNEB.}
	\label{fig:workflow}
\end{figure}

The workflow is illustrated in Fig.~\ref{fig:workflow}. Compared with standard NEB, there are mainly two modifications. First, before force evaluation, each structure is mapped into generalized coordinates via the cell filter described in the previous section. This allows atoms and lattice vectors to be treated on an equal footing during optimization. Second, after each NEB iteration, the image distribution is automatically adjusted by inserting or removing replicas based on local spacing in configuration space. This dynamic refinement is particularly useful when the initial guess is coarse or when localized structural changes occur only in part of the pathway.

We implemented CFNEB in two independent frameworks. The CPU-based version is built on the  python package Atomic Simulation Environment (ASE)\cite{hjorthlarsen_atomic_2017}, where we preserve the internal structure of the \texttt{BaseNEB} class and simply wrap each image with a \texttt{CFNEBCellFilter}. This design ensures full compatibility with existing ASE optimizers and calculators, allowing users to enable variable-cell NEB with minimal code modifications. he automatic interval adjustment is integrated as an observer in the optimizer, so that the path is continuously adjusted without user intervention.

The GPU-based implementation is built within the C++ molecular dynamics package GPUMD\cite{fan_neuroevolution_2021}. we managed to do most of vector and matrix manipulations on GPU. Aside from input/output operations, the structural information is always kept and updated on GPU, which keeps overhead negligible. On an NVIDIA GeForce 4090 GPU, we achieve a throughput of approximately $10^6$ atom·steps per second. This enables NEB calculations with on the order of 100 images and $10^5$ atoms to be propagated by one iteration in roughly 10 seconds, making large-scale transition-path sampling practically feasible. Limited by the available GPU memory, the maximal computational scale is about $10^8$ atom·images.

\section{Examples}
\subsection{$\beta$-$\lambda$ phase transition in \ce{Ti3O5}}
Liu et al. \cite{liu_layerbylayer_2024} reported that \ce{Ti3O5} transforms from the $\beta$ phase to the $\lambda$ phase via a layer-by-layer mechanism, identifying several metastable intermediates. The step-by-step nucleated pathway has a significant lower energy barrier than the concerted pathway. However, their simulations were limited to only making supercells in $a$ and $c$ directions, but not in $b$ direction. Using GPU-CFNEB combined with a NEP-trained force field (RMSE comparable to Ref. \cite{liu_layerbylayer_2024}, see Supplementary Material\cite{cfneb_supplementary}), we re-examined the transition with progressively larger supercells in $b$ direction.

\begin{figure}[t]
    \begin{center}
        \includegraphics[width=0.95\textwidth]{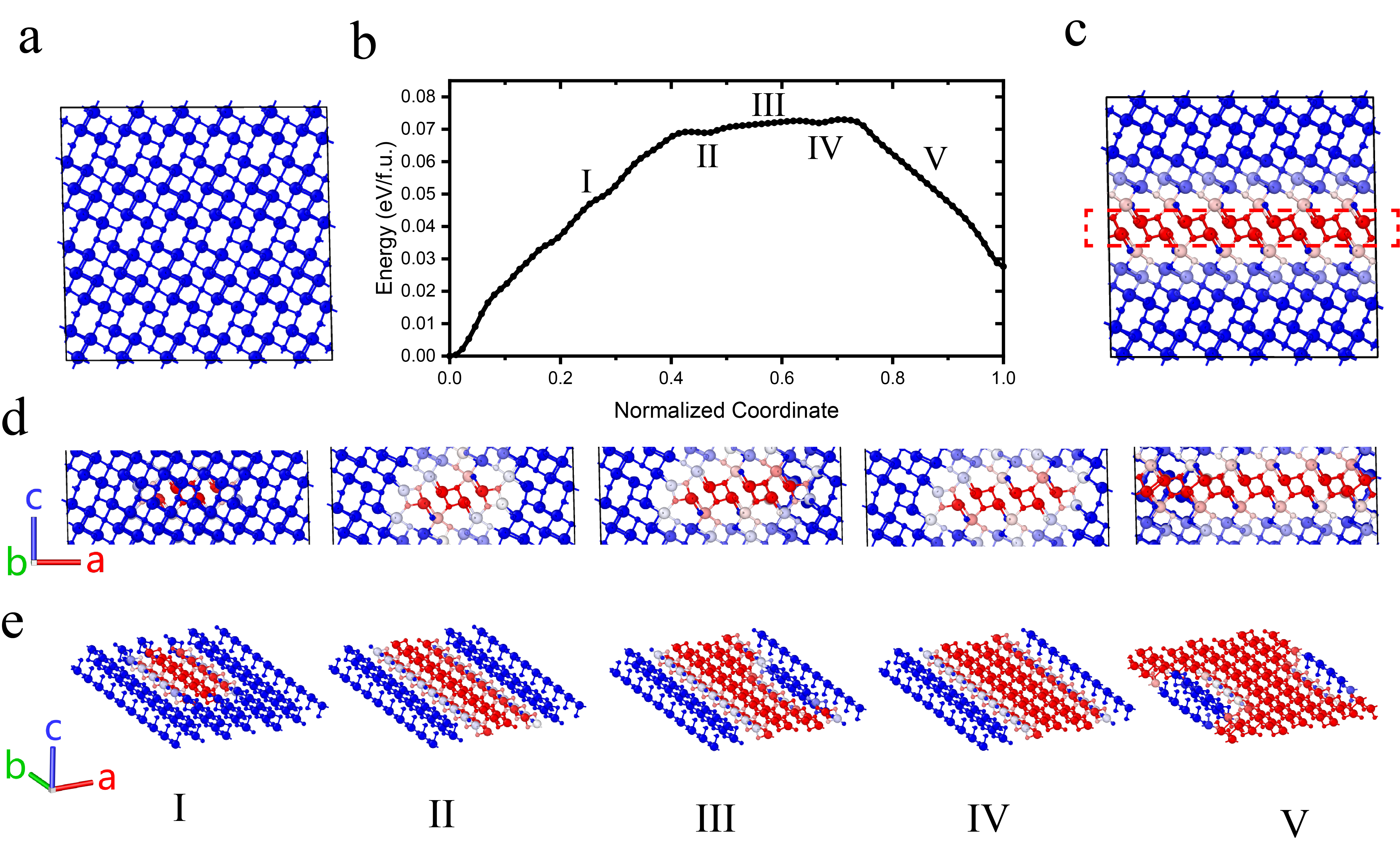}
    \end{center}
    \caption{The variable cell NEB pathway of \ce{Ti3O5} $\beta$-$\lambda$ phase transition. (a) the initial state of \ce{Ti3O5} $\beta$ phase. (b) The relative energy per formula with respect to the normalized coordinate. (c) the final state $\lambda$ phase. The atoms in blue color refer to the $\beta$ state and the ones in red refer to the $\lambda$ state. (d) is the front view of five intermediate states. (e) is the side view of five states and only the atoms sliced by the dash line rectangle in (c) are shown. II and IV are local minima. The large and small balls represent the Ti and O atoms and the colors are set by the normalized displacement magnitude of each atoms which represents the degree of location phase transition. The color legend is shown at the bottom. }
    \label{fig:tio}
\end{figure}

At small sizes, our results reproduce the previously reported concerted sliding of layers. However, once the simulation cell is extended by more than five additional units along the b-axis, the pathway spontaneously deviates from the synchronous motion and develops a helical or spiral-like displacement pattern. Figure~\ref{fig:tio} shows the transition path of the system with 10 supercells in $b$ direction. Instead of sliding uniformly, successive layers begin to rotate slightly out of phase, forming a screw-like propagation of the transition front. This emergent behavior—visible in Fig.~\ref{fig:tio}e—suggests that the true transition pathway may resemble a nucleation-and-growth process with local shear–rotation coupling, rather than a uniform layer shift. The preference for the spiral mode becomes stronger as the system size increases, indicating that such symmetry breaking is a finite-size effect inaccessible to traditional small-cell NEB calculations.

However, as mentioned in Liu's work, the final state $\lambda$ phase has a higher energy than the initial state $\beta$ phase without considering finite temperature effect, which results in infinitely growing total energy barrier as increasing the cell size. Nevertheless, the total energy barrier from intermediate meta-stable state IV to the final state is convergent. As we can see in the Fig.~\ref{fig:tio}b, the average energy barrier almost vanishes.

\subsection{Nucleation in graphite-to-cubic-diamond phase transition}
Diamond is widely recognized for its exceptional hardness, ultrawide electronic band gap, high thermal conductivity, and outstanding chemical stability \cite{dhaenens-johansson_synthesis_2022,field_mechanical_2012}. Despite the emergency of chemical vapor deposition (CVD), \cite{dhaenens-johansson_synthesis_2022} and plasma-assisted growth,\cite{chen_recent_2008} the high pressure high temperature (HPHT) method of compressing graphite is still the main method due to its multiple advantages, such as high crystal quality and low production cost.\cite{eaton-magana_observations_2017,guignard_review_2022} Nevertheless, a detailed atomistic understanding of the graphite-to-diamond transformation pathway remains incomplete \cite{khaliullin_nucleation_2011,zhu_revisited_2020,luo_atomistic_2024}.

\begin{figure}[t]
    \begin{center}
        \includegraphics[height=0.4\textwidth]{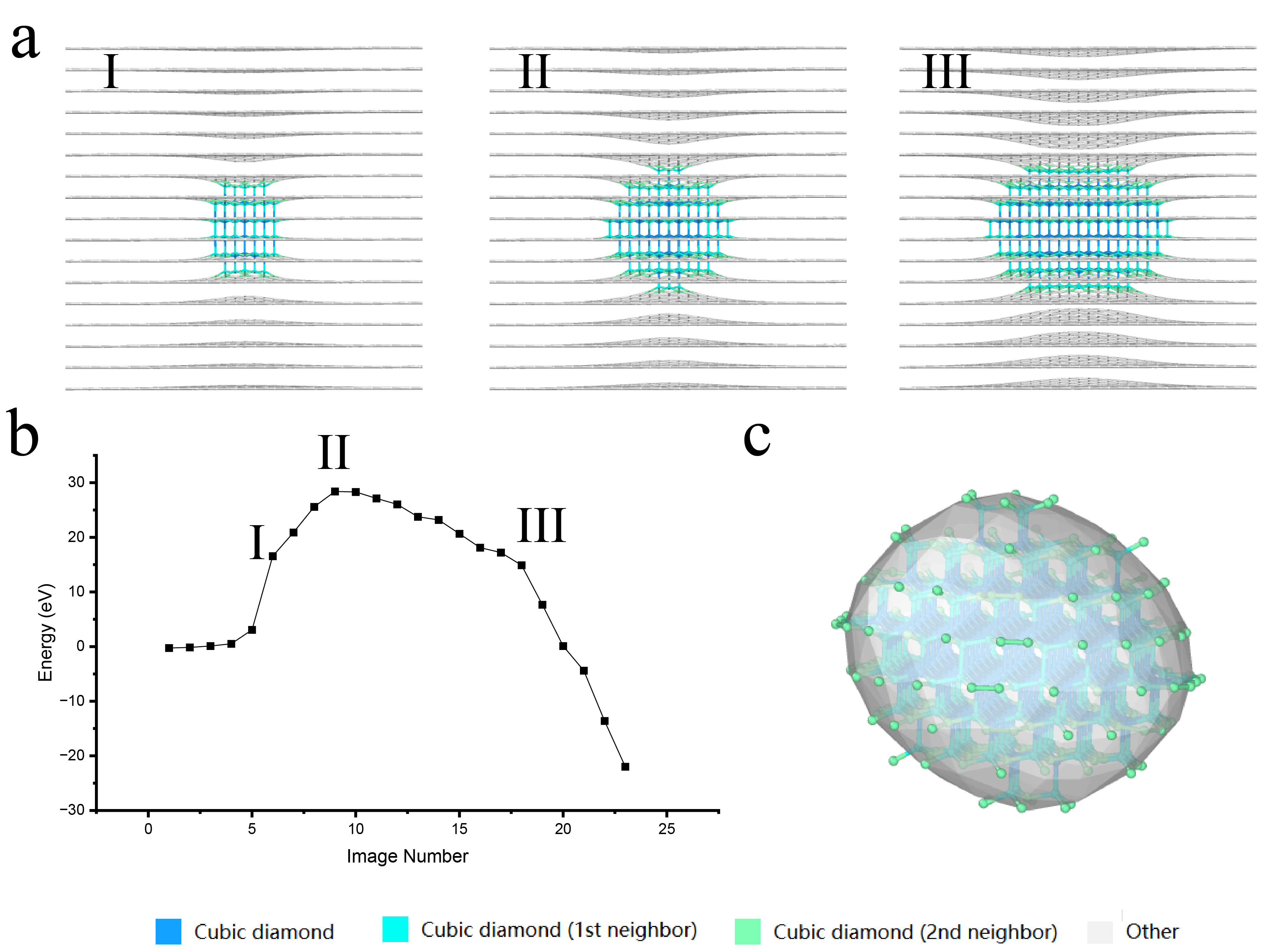}
    \end{center}
    \caption{The transition path of graphite-to-CD transition under 40GPa. (a) the intermediate states of the transition process. \textrm{II} is the saddle point. (b) the energy path from graphite to CD.  (c) the shape of nucleus extracted from state \textrm{II}. The atoms is colored by identify diamond structure modifier in OVITO.\cite{maras_global_2016}}
    \label{fig:gra2dia}
\end{figure}

In this work, we take a pristine graphite to cubic diamond (CD) transition as example to demonstrate the ability of GPU-CFNEB of probing this mechanism at realistic system sizes. Figure~\ref{fig:gra2dia}a shows the evolution of the structure along the reaction path. Initially, the graphite layers bend and form interlayer bonds, producing a small diamond-like cluster (Fig.~\ref{fig:gra2dia}a\textrm{I}).As the number of interlayer bonds increases, the nucleus expands and eventually reaches a critical size (Fig.~\ref{fig:gra2dia}a\textrm{II}), corresponding to the saddle point on the energy profile in Fig.~\ref{fig:gra2dia}b. The resulting nucleus is nearly spherical, but exhibits slight faceting along preferred crystallographic orientations, as shown in Fig.~\ref{fig:gra2dia}c.

Importantly, this nucleated pathway competes with the homogeneous “concerted buckling” mechanism often reported in smaller computational models. In sufficiently large supercells, CFNEB reveals that the nucleated mechanism becomes energetically preferred beyond a certain lateral dimension, consistent with recent ab initio molecular dynamics observations \cite{khaliullin_nucleation_2011,zhu_revisited_2020,luo_atomistic_2024}. This suggests that nucleation is not merely an alternative mechanism—but likely the dominant physical route under HPHT conditions, provided that system size is sufficient to accommodate localized fluctuations.

\section{Conclusion and Discussion}
\label{sec:conclusion}

We have developed a nudged elastic band framework capable of resolving transition pathways in systems much larger than those typically accessible to conventional implementations. By combining a deformation-based cell filtering scheme with adaptive image redistribution, CFNEB allows atomic and lattice degrees of freedom to be relaxed consistently while avoiding numerical artifacts related to rotational drift. The GPU implementation enables practical simulations involving $\mathcal{O}(10^7)$ atoms $\cdot$ images, with iteration times on the order of seconds on commodity hardware.

Applications to the $\beta$--$\lambda$ transition in \ce{Ti3O5} and the graphite-to-diamond transformation show that the method recovers previously reported concerted mechanisms in small systems while revealing deviations toward localized, nucleation-like pathways when size constraints are relaxed. This highlights the importance of cell size and boundary conditions in determining dominant transition modes and suggests that some reported concerted pathways may be finite-size artifacts rather than intrinsic mechanisms.

Two limitations should be noted. First, the present formulation treats only zero-temperature potential energy surfaces; incorporation of free-energy effects, e.g., via thermodynamic integration or umbrella sampling, would be required to assess kinetic preference under experimental conditions. Second, while adaptive image control improves numerical robustness, further development may be needed to fully automate image placement in highly anharmonic regions of configuration space.

Overall, CFNEB provides a practical and transferable approach for exploring large-scale transition mechanisms in solid-state systems. Its compatibility with both machine-learning force fields and AB initio accuracy levels suggests that it can be integrated into multiscale workflows for studying phase transformations, defect reactions, and mechanical deformation processes.

\section{Code available}
The python (ASE) version of NEB code is available in \url{https://gitlab.com/qhjia/vcneb}. The C++ CUDA (GPUMD) version is available in \url{https://gitlab.com/qhjia/GPUMD}. The projects are temporarily not public. Please contact the authors if you want to get access.

\section{Acknowledgment}
We gratefully acknowledge the financial support from the National Key R\&D Program of China (grant no. 2022YFA1403201), the National Natural Science Foundation of China (grant number. T2495231, 12125404, 123B2049), the Basic Research Program of Jiangsu (Grant BK20233001, BK20241253), the Jiangsu Funding Program for Excellent Postdoctoral Talent (2024ZB002, 2024ZB075), the Postdoctoral Fellowship Program of CPSF (Grant GZC20240695), the AI \& AI for Science program of Nanjing University, and the Fundamental Research Funds for the Central Universities. The calculations were carried out using supercomputers at the High Performance Computing Center of Collaborative Innovation Center of Advanced Microstructures, the high-performance supercomputing center of Nanjing University

\bibliography{cfneb}

\end{document}